\documentclass[fleqn,usenatbib]{mnras}

\usepackage{newtxtext,newtxmath}
\usepackage[T1]{fontenc}

\DeclareRobustCommand{\VAN}[3]{#2}
\let\VANthebibliography\thebibliography
\def\thebibliography{\DeclareRobustCommand{\VAN}[3]{##3}\VANthebibliography}

\usepackage{graphicx}	% Including figure files
\usepackage{amsmath}	% Advanced maths commands
\usepackage{amssymb}	% Extra maths symbols

\usepackage{multirow}
\newcommand{\hesssource}{HESS J1841-055} 
\newcommand{\fgl}{4FGL J1840.9-0532e}
\newcommand{\thrirdfhl}{3FHL J1840.9-0532e}

\title[MAGIC observation of \hesssource]{Studying the nature of the unidentified gamma-ray source \hesssource\ with the MAGIC telescopes}

\author[Acciari et al.]{
  \parbox{\linewidth}{
V.~A.~Acciari$^{1}$,
S.~Ansoldi$^{2}$,
L.~A.~Antonelli$^{3}$,
A.~Arbet Engels$^{4}$,
K.~Asano$^{5}$,
D.~Baack$^{6}$,
A.~Babi\'c$^{7}$,
A.~Baquero$^{8}$,
U.~Barres de Almeida$^{9}$,
J.~A.~Barrio$^{8}$
% ...
(MAGIC Collaboration)
}
}

\date{Accepted XXX. Received YYY; in original form ZZZ}

\pubyear{}

\begin{document}
\label{firstpage}
\pagerange{\pageref{firstpage}--\pageref{lastpage}}
\maketitle

% Abstract of the paper
\begin{abstract}
We investigate the physical nature and origin of the gamma-ray emission from the extended source \hesssource\ observed at TeV and GeV energies.
We observed \hesssource\ at TeV energies for a total effective time of 43 hours with the MAGIC telescopes, in 2012 and 2013. Additionally, we analysed the GeV counterpart making use of about 10 years of \textit{Fermi}-LAT data. Using both \textit{Fermi}-LAT and MAGIC, we study both the spectral and energy-dependent morphology of the source for almost four decades of energy. The origin of the gamma-ray emission from this region  is investigated
   using multi-waveband information on sources present in this region, suggested to be associated with this unidentified gamma-ray source. We find that the extended emission at GeV-TeV energies is best described by more than one source model. We also perform the first energy-dependent analysis of the HESS J1841-055 region at GeV-TeV. We find that the emission at lower energies comes from a diffuse or extended component, while the major contribution of gamma rays above 1 TeV arises from the southern part of the source.  Moreover, we find that a significant curvature is present in the combined observed spectrum of MAGIC and \textit{Fermi}-LAT. 
   The first multi-wavelength spectral energy distribution of this unidentified source shows that the emission at GeV--TeV energies can be well explained with both leptonic and hadronic models. For the leptonic scenario, bremsstrahlung is the dominant emission compared to inverse Compton. On the other hand, for the hadronic model, gamma-ray resulting from the decay of neutral pions ($\pi^0$) can explain the observed spectrum. The presence of dense molecular clouds overlapping with \hesssource\  makes both bremsstrahlung and $\pi^0$-decay processes the dominant emission mechanisms for the source.
\end{abstract}


%-------------------------------------------------------------------

%%%%%%%%%%%%%%%%%%%%%%%%%%%%%%%%%%%%%%%%%%%%%%%
%%%%%%%%%%%%%%%%%%%%%%%%%%%%%%%%%%%%%%%%%%%%%%%
\section{Introduction}
The unidentified gamma-ray source \hesssource\ was first discovered at TeV energies in 2007 by the High Energy Stereoscopic System (H.E.S.S.) during the Galactic plane survey \citep{Aharonian_2008_AA}. The observed emission was reported as extended with an elliptical extension of 0.41$^\circ$ and 0.25$^\circ$ along the semi-major and semi-minor axes, respectively and centered at Right Ascension (RA):18$^h$40$^m$55$^s$ and declination (Dec): 5$^\circ$33$'$00$''$ with a position angle 39$^\circ$ relative to the RA axis. 
\hesssource\ was detected with a statistical significance of 10.7$\sigma$ and a flux of (12.8$\pm$1.3) $\times$ 10$^{-12}$ cm$^{-2}$ s$^{-1}$ between 0.54 and 80 TeV. The spectrum is best described by a powerlaw with a spectral index of $\rm 2.41 \pm 0.1_{stat} \pm 0.2_{sys}$. These results are compatible with the recent results reported by H.E.S.S. collaboration \citep{HGPS_2018}. 
Using the ARGO-YBJ experiment for energies above 0.9 TeV, \citet{Bartoli_2013ApJ} reported a similar extension as seen be the H.E.S.S collaboration but a 3 times larger flux due to differing background estimation techniques between the experiments.
This region was further investigated by the HAWC observatory also at TeV energies. The source 1HWC J1838-060, from the First HAWC Catalog, was detected at 6.1$\sigma$ post-trial significance. It is located in the middle of \hesssource\ and another known TeV source, HESS J1837--069 \citep{Abeysekara_2016_ApJ}. This detection by HAWC was found to be overlapping with the extension of HESS J1841--055, and the differential flux normalization was compatible with the one reported by the H.E.S.S. collaboration. The second HAWC Catalog also revealed a source, 2 HWC J1837-065 which was likely to be associated to \hesssource \citep{2nd_HAWC_2017ApJ...843...40A}. Its spectral index varies from -2.90$\pm$0.04 for a point-like emission to -2.66$\pm$0.03 for a 2$^{\circ}$ radius.

This region was further investigated at other wavelengths to search for possible counterparts. Although no confirmed counterparts of the TeV source \hesssource\ at lower energies are known, several possible associations have been suggested. The emission from \hesssource\ may be due to either a single extended source or several unresolved sources. \citet{Sguera_2009_ApJ}, making use of INTEGRAL data, proposed as counterpart the unidentified transient source 3EG J1837--0423, which was likely to be associated to the Supergiant Fast X-ray Transient (SFXT) AX J1841.0--0536. At X-ray energies, observations of this extended region were done with \textit{SUZAKU} and an X-ray source was discovered \citep{Nobukawa_2015_AdSpR}. The detection of two separate extended sources (FGES J1839.4-0554 and FGES J1841.4-0514) was also reported in this region at energies above 10 GeV using data from the Fermi-Large Area Telescope \citep[LAT,][]{Ackermann_FGES_2017ApJ,3FHL_2017ApJS..232...18A}. Some potential sources at different wavelengths suggested to be associated with \hesssource\ is discussed later in detail. 

In this paper, we study this complex region using dedicated observations with the MAGIC telescopes at TeV energies. We also explore the GeV counterpart making use of 10-year data of \textit{Fermi}-LAT. We finally model the GeV-TeV emission to unveil the dominant gamma-ray emission mechanisms at work. The potential counterparts at other frequencies are also investigated. The low energy threshold of MAGIC, which allows to overlap with \textit{Fermi}-LAT in the GeV domain, combined with the MAGIC capabilities of reaching several TeV, make the MAGIC telescopes a suitable instrument to study this region within a broad energy range. The combination of both MAGIC and \textit{Fermi}-LAT allows spectral studies of this complex region for almost four decades in energy.

The paper is organized as follows: in section \ref{sec:data_analysis}, we describe the detailed analyses of the MAGIC and \textit{Fermi}-LAT data. The results are discussed in section \ref{sec:results}. Potential counterparts are proposed in section \ref{sec:conterparts}. The multiwaveband modelling of the source is explained in section \ref{sec:Modelling}. Finally, we summarize and conclude in section \ref{sec:Discussion}.       

%%%%%%%%%%%%%%%%%%%%%%%%%%%%%%%%%%%%%%%
%%%%%%%%%%%%%%%%%%%%%%%%%%%%%%%%%%%%%%%
\section{Observations and data reduction} \label{sec:data_analysis}
%+++++++++++++++++++++++++++++++
%+++++++++++++++++++++++++++++++
\subsection{MAGIC}
Very-High-Energy (VHE, E> 100 GeV) gamma-ray observations of \hesssource\ {are} performed using the MAGIC telescopes. MAGIC consists of two 17 m diameter Imaging Atmospheric Cherenkov Telescopes (IACTs) located at the Observatory of Roque de los Muchachos (28$^\circ$.8 N, 17$^\circ$.9 W, 2200 m above the sea level) {on} the Canary Island of La Palma, Spain. The  energy threshold of the MAGIC stereoscopic system is about 50 GeV, and it is able to detect $\sim$ 0.6\% of the Crab Nebula flux above 250 GeV at 5$\sigma$ significance in 50 hours of observations at small ($<$30$^\circ$) zenith angles \citep{MAGIC_performance_2012}. 
\hesssource\ was observed between April 2012 and August 2013, for a total of about 43 hours, at zenith angles between 5$^\circ$ and 50$^\circ$,resulting in an energy threshold for this analysis of  $\sim$150 GeV. To estimate the background simultaneously {with the source data}, the observations are performed in the so-called wobble-mode \citep{Fomin_1994} at two symmetrical positions, with the source located $0^\circ.55$ off-axis from the center of the camera. After quality cuts, which account for hardware problems, unusual rates, and bad atmospheric conditions, $\sim$34 hours of high-quality, dark-time data are selected for further analysis.
The analysis of the MAGIC data is performed using the standard MAGIC Analysis and Reconstruction Software  \citep[Mars;][]{Moralejo_2009,Zanin_2013} and standard analysis procedure. 

Given the extension of the source and the possibility of contamination from other nearby sources, we study the region using an iterative maximum likelihood method included in the \textit{Skyprism} package \citep{Vovk_2018A&A}. \textit{Skyprism} has specifically been developed to perform 2D fitting of IACTs data and has been optimised for MAGIC data. This set of tools compute the instrument response function (IRF) and perform a simultaneous maximum likelihood fit of source models of arbitrary morphology to the sky images. With \textit{Skyprism} it is then possible to analyse MAGIC data of extended sources of arbitrary morphology and multiple, overlapping sources.

{We compute the event count map, the background map, and the instrument response functions which includes Point Spread Function (PSF), energy migration matrix, and exposure map. We use the "exclusion map" method for generating the background map excluding a circular region of $0^\circ.5$ around the center of the \hesssource\ (RA = $280^{\circ}.23$, Dec=$-5^\circ.55$) and a circular region of $0^\circ.3$ around a bright spot at the southern edge of the camera (RA = $279^{\circ}.4$, Dec=$-6^\circ.45$). A user-defined source model (2D Gaussian) is used to fit the measured event maps for maximizing the log-likelihood estimate.} To calculate the individual spectral parameters of the sources obtained from the modelling of the region, we use {the} maximum log-likelihood method, as defined in \cite{Vovk_2018A&A}, assuming a powerlaw model for the source at the pivot energy 1 TeV (energy at which the uncertainty in the normalisation is minimum). The observations for this work are performed at low and medium zenith angles (Z<50$^\circ$). Given the very high signal-to-noise ratio ($>$0.4), the systematic uncertainties can be considered similar to those reported in \citet{MAGIC_performance_2012}, defined as 12\% in the integral flux for stereoscopic observations.

%+++++++++++++++++++++++++++++++
%+++++++++++++++++++++++++++++++
\subsection{\textit{Fermi}-LAT}\label{sec:fermi-lat_analysis}
The {\it Large Area Telescope} (LAT) onboard the \textit {Fermi Gamma Ray Space Telescope}  allows for the detection of gamma rays from 30 MeV to $>$ 500 GeV with its large effective area and wide field of view \citep{Atwood2009}. In our analysis, we select nearly ten years (i.e., from 2008 September 1 to  2017 May 5) of Pass 8 SOURCE class (P8R3) LAT events in the reconstructed energy of about {10 GeV} to 1 TeV within a 15$^{\circ}$ region of interest (ROI) around the fourth \textit{Fermi}-LAT catalog source \fgl\ ({associated to} \thrirdfhl). {TeV source} \hesssource\ {is associated with the \textit{Fermi-}LAT source \fgl}.  The Fermi Science Tools (FST) analysis package\footnote{\url{fermi.gsfc.nasa.gov/ssc/data/analysis/software/}} version \texttt{v11r5p3} and the \texttt{P8R3$_{-}$SOURCE$_{-}\!\!$V2} IRFs are used for the analysis. We also use the python-based package \texttt{Fermipy} (version 0.17.4\footnote{\url{https://fermipy.readthedocs.io/en/latest/}}) to facilitate the analysis of data with the FSTs. {We} select photons of energies greater than {10 GeV} with arrival direction within 105$^\circ$ from local zenith to remove contamination from the Earth's emission. The PASS 8 source class allows for the use of {four} different \textit{event types} which are based on the event-by-event quality of reconstructed direction (PSF) and energy. {Hence, the data is separated into these event types to optimize selection of events based on the quality of reconstruction of direction of incoming photons and energy}. The Galactic diffuse emission is modeled with the standard \textit{Fermi}-LAT diffuse emission model \textit{(gll\_iem\_v07.fits)}. The isotropic emission from extragalactic radiation and residual background models \textit{(iso\_P8R3\_SOURCE\_V2\_PSF[0/1/2/3]\_v1.txt)} are also used corresponding to four event types.

 We first start with a baseline sky model which includes all {4FGL point and extended sources} within the ROI listed in the 4FGL catalogue\footnote{\url{https://fermi.gsfc.nasa.gov/ssc/data/access/lat/8yr_catalog/gll_psc_v19.fit}}\citep{4FGL_catalog_2020ApJS..247...33A}. The extended source \fgl\ ({associated to} \thrirdfhl) is our source of interest in the ROI which is associated with two sources from the Fermi Galactic Extended Source Catalog (FGES) \citep{Ackermann_FGES_2017ApJ} and it is included in the model. Initially, we use  {the} baseline model {to optimize parameters of the sources} by fitting their flux and spectral parameters. After the initial optimization, we remove all sources for which the values of the predicted number of counts in the model, \texttt{Npred}, are less than 2.0 and we free spectral shapes and normalizations for all the sources which lie within 3$^\circ$ from the center of the ROI.  The isotropic diffused background model is fixed to the value obtained after the first optimization of the ROI but the diffuse Galactic model is kept free for all different configurations or models discussed below. Then we use the binned maximum likelihood method to estimate the best-fit model parameters using a 15$^\circ \times 15^\circ$ square region centered on {\fgl}\ with {a spatial bin-size of 0$^\circ$.06} and 10 equally spaced energy bins per decade of energy. We then relocate the source of interest using the maximum likelihood method to find the best source position. As the next step, we use an iterative maximum likelihood-based source finding algorithm to identify new point sources within 0.5 deg from the centre of the ROI. The algorithm finds point sources within this ROI with test-statistics\footnote{The Test Statistic (TS) of a source is evaluated using a likelihood ratio test defined as TS = $\rm -2 log({\mathcal{L}_1 \over \mathcal{L}_0})$, where $\mathcal{L}_0$ and $\mathcal{L}_1$ are the likelihoods of the background model without the source (null hypothesis) and the hypothesis being tested (source plus background), respectively. The detection significance is approximately the square root of the TS.}, TS $>$ 16. We continue searching for new sources until all the point sources are added to the baseline model. Following this, we remove all the sources with TS $ < $ 16 from the ROI and perform the maximum likelihood method for the best-fit model parameters.

%%%%%%%%%%%%%%%%%%%%%%%%%%%%%%%%%%%%%%%
%%%%%%%%%%%%%%%%%%%%%%%%%%%%%%%%%%%%%%%
\section{Results} \label{sec:results}

%+++++++++++++++++++++++++++++++
%+++++++++++++++++++++++++++++++
\subsection{MAGIC}\label{sec:MAGIC_results}
\subsubsection{Morphology}
\begin{figure*}
\begin{tabular}{ccc}
\centering
\includegraphics[width=0.33\textwidth]{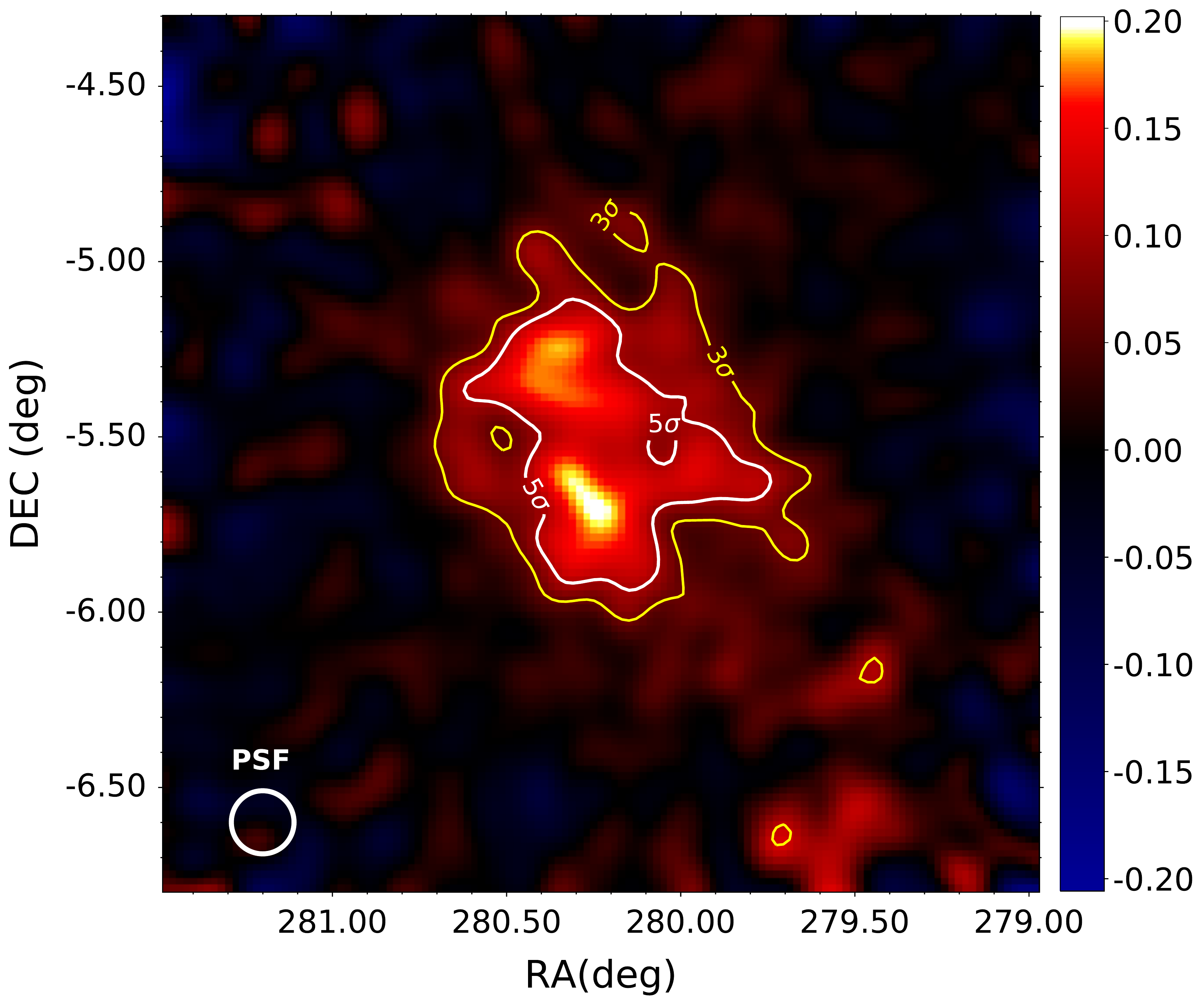}
\includegraphics[width=0.33\textwidth]{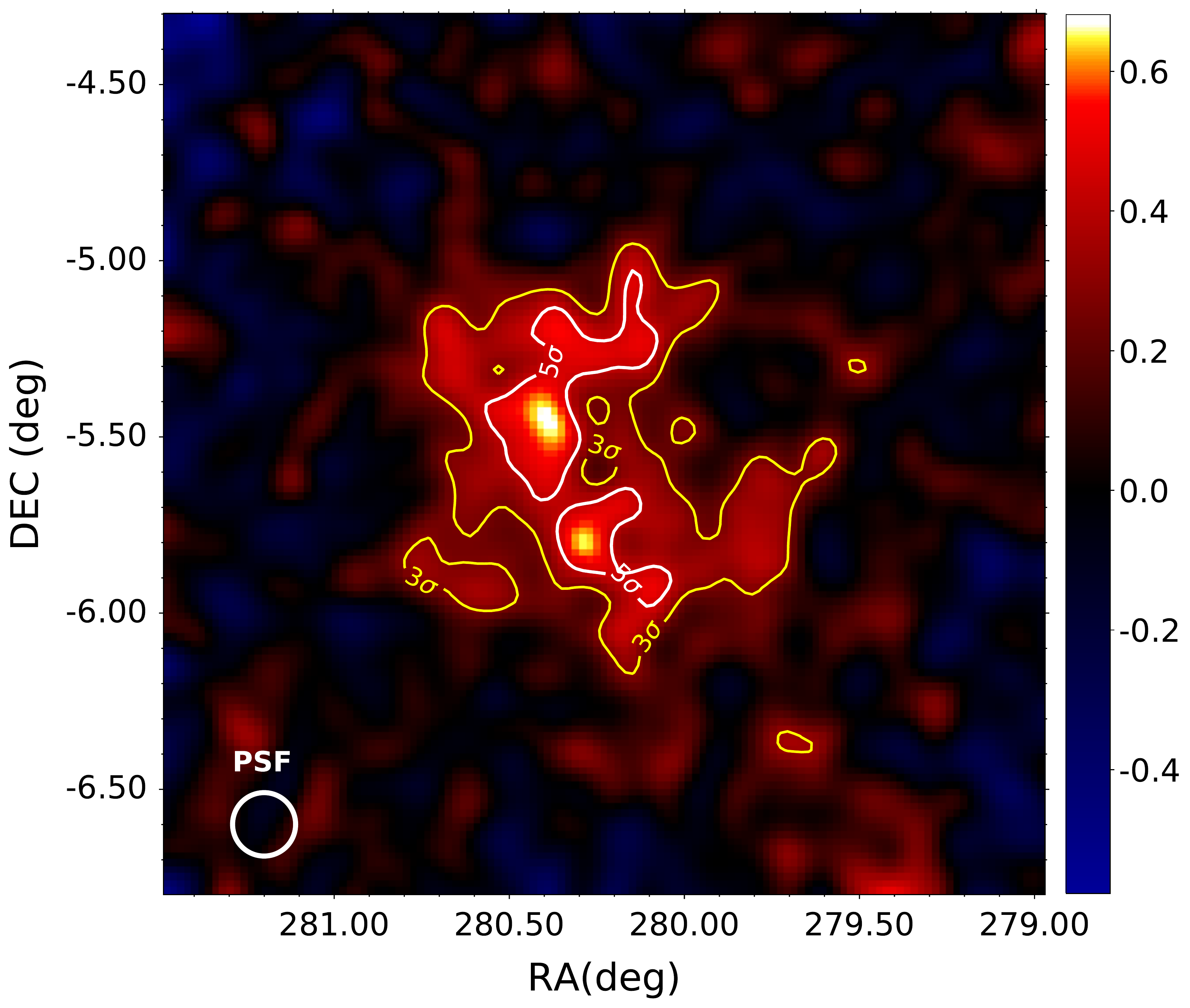}
\includegraphics[width=0.34\textwidth]{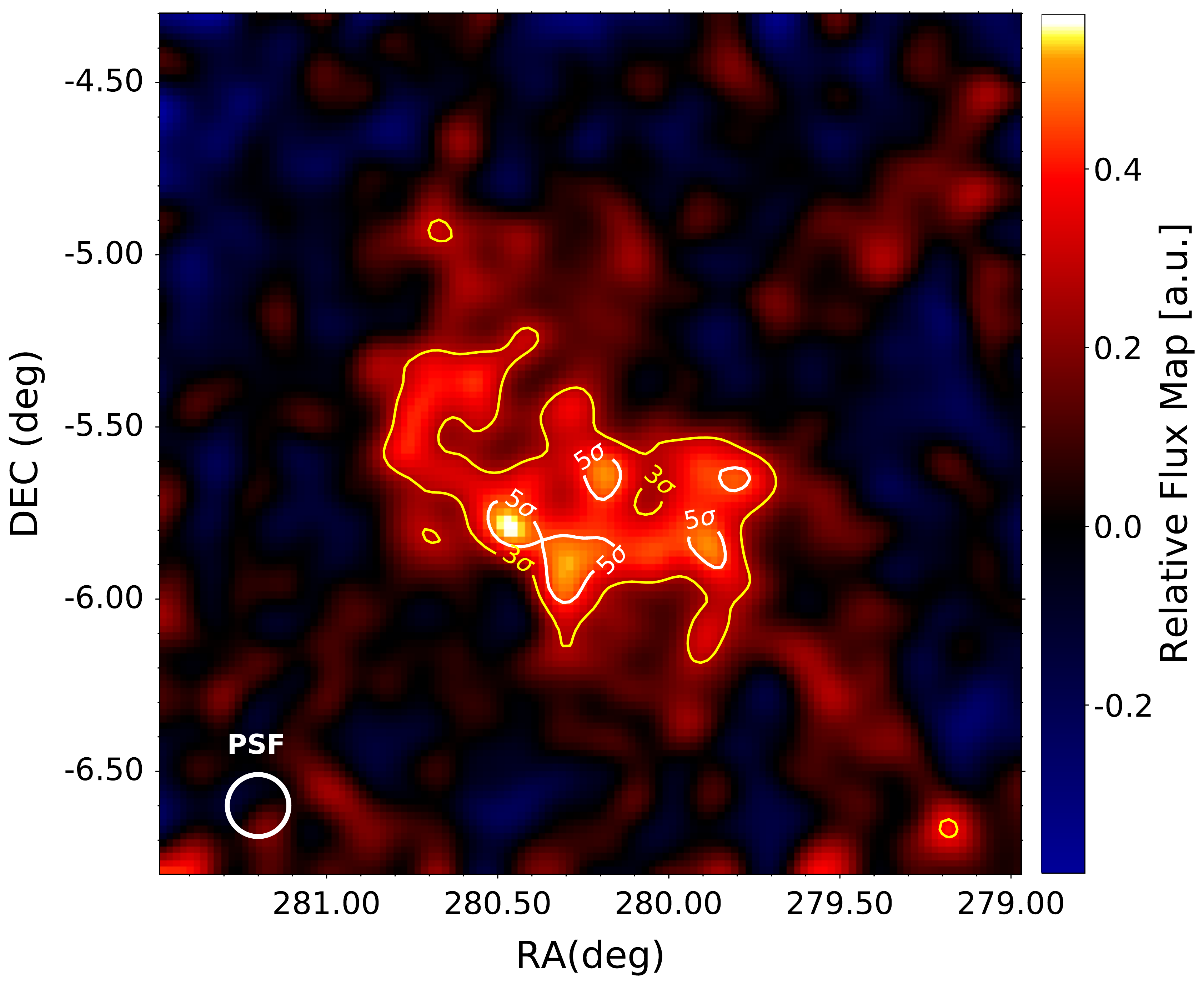}
\end{tabular}
\caption{Energy-dependent gamma-ray relative flux maps with 3$\sigma$ (yellow) and 5$\sigma$ (white) contour levels of the extended source \hesssource\ as seen by MAGIC.  The energy ranges covered are LE (50--500 GeV), {ME} (500 GeV -- 1 TeV) and HE ($>$ 1 TeV), shown in the three panels, from left to right, respectively.}\label{fig:skymaps}
\end{figure*}

In order to study the energy-dependent morphology of the extended source \hesssource, we produce skymaps for different energy ranges using \textit{Skyprism}. Fig. 
\ref{fig:skymaps} shows the relative flux skymaps, with 3$\sigma$ and 5$\sigma$ contours extracted from the TS map, produced for energies 50 -- 500 GeV (low energy, LE, map), 500 GeV -- 1 TeV  ({medium energy, ME}, map) and $>$1 TeV (high energy, HE, map), respectively. {The relative flux is defined as the excess events divided by the background events.} To calculate the extension of the source in each of the energy ranges, we consider a radially symmetrical two-dimensional (2D) Gaussian shape. The 1$\sigma$ standard deviation of the Gaussian is considered to be the extension or radius of the source. The radius is kept as a free parameter and is allowed to change by 0$^\circ$.01 over a range of 0$^\circ$.1  to 0$^\circ$.6 during the maximum likelihood fitting. Moreover, we simultaneously keep changing the origin of the Gaussian by changing RA and Dec by 0$^\circ$.01 for both of them over a range of 1$^\circ$. The best-fit locations along with the extensions of the source for different energy ranges are shown in Table \ref{tab:fit_params_extension}. The extension of the source at these three energy ranges appears to be the same, however, the overall detection significance of the extended emission reduces at higher energies, revealing only a few hotspots in the southern part of the source  (see Fig. \ref{fig:skymaps}). The fitted extension is the same (within errors) in the whole energy range (see Table \ref{tab:fit_params_extension}). MAGIC observations show that the source has an extension compatible with that measured by H.E.S.S. collaboration at TeV energies. It is also evident from the different maps that the extended region shows  several bright {hotspots} with a significance of more than 5$\sigma$. Many bright highly-significant spots are detected at LE and {ME} energies, while they mostly disappear at HE. These hotspots hint the presence of multiple sources in the region. It also indicates that the most significant emission at higher energies is coming from the southern part of the region.

 \begin{table}
     \centering
      \caption{Best-fit parameters of the extension of the source measured by MAGIC considering a symmetrical 2D Gaussian model.}
     \begin{tabular}{c|c|c|c}
     \hline 
     \hline
         Energy range    & RA ($^\circ$) & Dec ($^\circ$) & Extension ($^\circ$)  \\
         \hline
          50 - 500 GeV        & 280.27$^{+0.03}_{-0.04}$ & -5.59$^{+0.02}_{-0.03}$ & 0.39$^{+0.21}_{-0.15}$ \\
          500 -1000 GeV       & 280.29$^{+0.01}_{-0.04}$ & -5.58$^{+0.01}_{-0.05}$ & 0.42$^{+0.04}_{-0.19}$ \\
          > 1000 GeV      & 280.29$^{+0.01}_{-0.04}$ & -5.70$^{+0.01}_{-0.05}$ & 0.45$^{+0.04}_{-0.04}$ \\ 
            \hline
            \hline
    \end{tabular}
        \label{tab:fit_params_extension}
 \end{table}

{As discussed above, the extended source \hesssource\ may potentially consist of multiple sources.  To check this, we consider three different source models covering the full energy range, i.e., energies from 50 GeV to above 1 TeV. We first consider a single-source model where the extended source is considered to be a 2D elliptical Gaussian. We leave the peak position, extension along X and Y direction and angle w.r.t. the X direction free while maximizing the likelihood value of the fit. For the second model, we replace the single-source model with two sources which are modelled as 2D circular Gaussian. The peak position and radius (1$\sigma$ standard deviation) of the two sources are free parameters of the model.  Finally, for the third option, we model the entire source region considering three different sources, one with elliptical disk model and the other two with Gaussian models. The results of the maximum likelihood values are given in Table \ref{tab:models_multi_source}. It is found that both two-source model and three-source models are better than a single-source model. The improvement of the two-source model w.r.t. to the one-source model is given by TS = 9.4 for 1 additional degree of freedom (d.o.f.), which corresponds to an improvement at 3$\sigma$. The improvement of the three-source model w.r.t. to the one-source model is given by test-statistics of 16.2 for additional 5 d.o.f which corresponds to an improvement of 2.7$\sigma$. This hints that the \hesssource\ region is better modelled by multiple sources. The parameters of the best-fit models are shown in Table \ref{tab:models_best-fit_multi_source}. }

\begin{table}
    \centering
     \caption{Significance of the multiple sources at TeV energies for different spatial source models.}
    \begin{tabular}{c|c|c}
    \hline
    \hline
        Spatial model      & $\Delta \log\mathcal{L}$\footnotemark & d.o.f \\
        \hline
       One Elliptical Gaussian model             &  0.0          &  5    \\
       Two Gaussian models                       &  4.7          &  6    \\
       Two Gaussian + one elliptical disk models &  8.1          &  11    \\
      \hline 
    \end{tabular}
    \label{tab:models_multi_source}
   \footnotetext{}{$^5$Calculated w.r.t. \textit{one-source} model}
\end{table}
\setcounter{footnote}{0}

\begin{table*}
    \centering
     \caption{Parameters of the best-fit single and multi-source models. The ext$_x$ and ext$_y$ are extension of the models along X and Y direction respectively. For elliptical Gaussian, they are standard deviation, whereas for elliptical disk model they correspond to semi-major and semi-minor axis respectively. }
    \begin{tabular}{c|c|c|c|c|c|c}
    \hline
    \hline
        Spatial model      & sources & Ra[$^\circ$] & Dec[$^\circ$] & ext$_x[^\circ$] & ext$_y[^\circ$] & $\theta[^\circ]$ \\
                          
        \hline
       One Elliptical Gaussian model  & source1 &  280.21 $\pm$ 0.02 & -5.57 $\pm$ 0.02  & 0.31 $\pm$ 0.02 & 0.46 $\pm$ 0.03 & 149.5 $\pm$ 5.7    \\
       \hline
       \multirow{3}{*}{Two Gaussian models}     &  source1 & 280.28 $\pm$	0.02 & -5.48 $\pm$0.03          & 0.34 $\pm$	0.02 & - & -    \\
        & source2  & 279.80 $\pm$	0.22 & -6.12 $\pm$ 0.09 & 0.37 $\pm$ 0.06 & -   & - \\
        \hline
        \multirow{3}{*}{Two Gaussian + one elliptical disk models}& source1 & 280.29 $\pm$0.05 & 0.32 $\pm$	0.05  & -	          &  -    \\
        & source2    &  279.78	$\pm$ 0.30  & -6.11 $\pm$	0.30 &  0.37 $\pm$	0.22 & - & -\\
        & source3    &  280.18 $\pm$	0.07    &  -5.45	$\pm$ 0.04 & 0.32 $\pm$	0.02 & 0.60 $\pm$	0.03 & 122.08 $\pm$	2.20	 \\
      \hline 
    \end{tabular}
    \label{tab:models_best-fit_multi_source}
\end{table*}
\renewcommand*{\thefootnote}{\arabic{footnote}}

%+++++++++++++++++++++++++++++++
%+++++++++++++++++++++++++++++++
\subsubsection{Spectrum}\label{sec:magic_results_spectrum}
\begin{figure}
\includegraphics[width=0.45\textwidth]{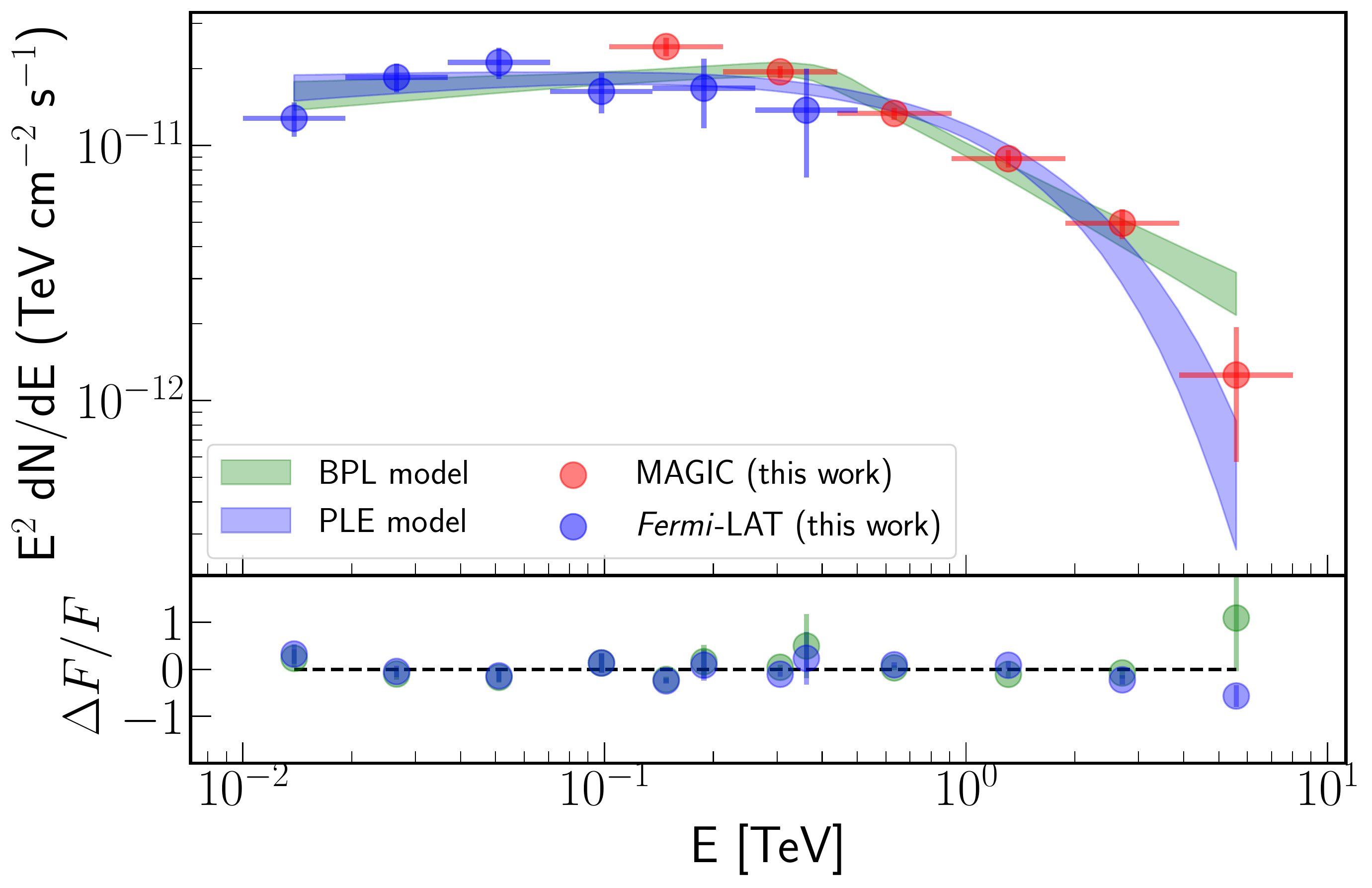}
\caption{The spectral energy distribution (SED) of the extended source \hesssource. The MAGIC energy fluxes are shown for energy above 100 GeV (red) whereas \textit{Fermi}-LAT energy fluxes are obtained for energy above 10 GeV (blue). The combined fit of the \textit{Fermi}-LAT  and MAGIC SEDs is best described by either a BPL model (green shaded region) or a PLE (blue shaded region) model. }\label{fig:sed_magic}
\end{figure}

The spectral energy distribution (SED) is calculated in the energy range of 50 GeV to $>$ 1 TeV, using the \textit{Skyprism} package. We consider the extended 2D Gaussian template with the extension 0.4$^\circ$ at the position of the \hesssource\ and an isotropic background. The assumed spectrum of the source is considered to follow a simple powerlaw (PL) model which is defined as follows:
$$\rm PL: {dN \over dE }= N_0 \left({E \over E_0}\right)^{-\alpha},$$
\noindent where $N_o$ and $\alpha$ are parameters of the model. The best-fit spectral parameters are $\rm N_o =(9.43 \pm 0.29) \times 10^{-12} \rm TeV^{-1}~ cm^{-2}~ s^{-1}, \alpha = 2.57 \pm 0.05$. The gamma-ray flux above 50 GeV is, $\rm F (>50 ~GeV) = 2.23 \times 10^{-10} ~cm^{-2}~ s^{-1}$. The SED measured by MAGIC is plotted in Fig. \ref{fig:sed_magic}.

Although morphology studies reveal that the emission region can be modeled better with more than one source, we cannot make any robust estimate on the number of distinct sources due to limitations of the software tool. Hence we do not provide high-quality SEDs associated with these sources.

%+++++++++++++++++++++++++++++++
%+++++++++++++++++++++++++++++++
\subsection{\textit{Fermi}-LAT}\label{sec:FermiLAT_ data}
 
%+++++++++++++++++++++++++++++++
%+++++++++++++++++++++++++++++++
\subsubsection{Morphology}\label{sec:fermi_morphology}
\begin{figure}
\includegraphics[width=0.45 \textwidth]{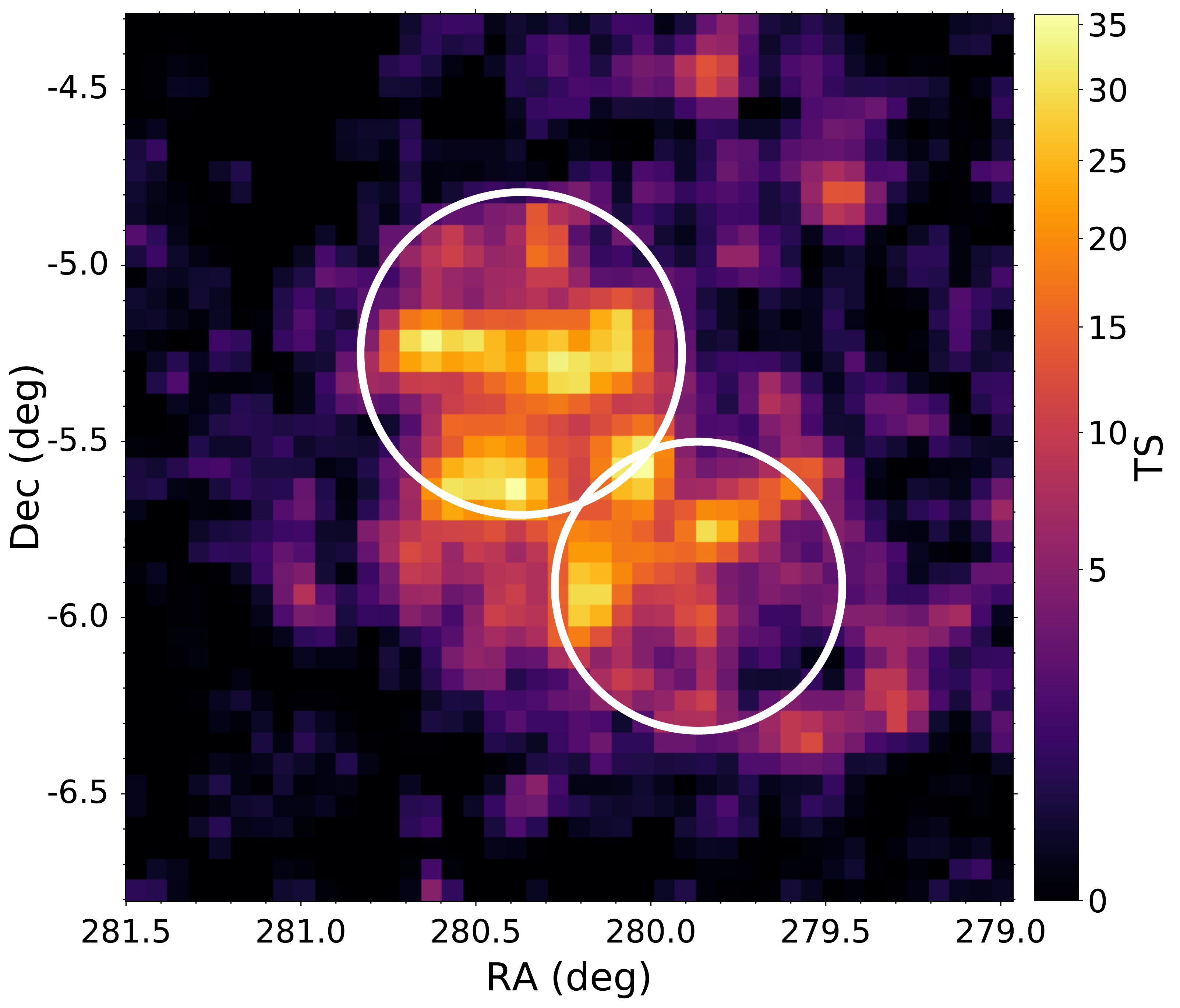}
\caption{\textit{Fermi}-LAT TS map (in Galactic coordinates) for the energy range from 10 GeV to 1 TeV. The two FGES sources are also shown (white circles).}\label{fig:fermi_skymap}
\end{figure}

For the morphological analysis of the source, photons with energy above 10 GeV up to 1 TeV are considered to reduce the contamination from nearby pulsars within the ROI. With the baseline model, as discussed in Section \ref{sec:fermi-lat_analysis}, we perform the binned maximum likelihood analysis on \fgl\ and find the best-fit model parameters.
To estimate the size of \fgl, we calculate the TS of the extension ($\rm TS_{ext}$) parameter, which is the likelihood ratio of the likelihood for being a point-like
source ($\rm L_{pt}$) to a likelihood for an assumed extension ($\rm L_{\rm ext}$), $\rm TS_{\rm ext} = 2log(\rm L_{\rm ext}/L_{pt}$). 
In order to test the extension of the source of interest, a radially symmetric Gaussian is considered and we vary its sigma from 0$^\circ$.01 to  1$^\circ$.5 in steps of 0$^\circ$.1. We also simultaneously leave the location of the center of the source free within 1$\sigma$ extension of the Gaussian. We find that the source extension is 0$^\circ.64 \pm 0^\circ.11$ with the $\rm T_{ext}$ = 264 which corresponds to a significance of about 16$\sigma$. We also consider a radial disk model and found that the source extension is  0$^\circ.60 \pm 0^\circ.11$ with the $\rm T_{ext}$ = 224, which corresponds to a significance of about 15$\sigma$. However, the log-likelihood {is} maximum for the Gaussian model,  {hence it will be} considered as the preferred model for {\fgl}. The resulting \textit{Fermi}-LAT TS map above between 10 GeV and 1 TeV is shown on Fig. \ref{fig:fermi_skymap}.

%+++++++++++++++++++++++++++++++
%+++++++++++++++++++++++++++++++
\subsubsection{Spectrum}\label{sec:spectrum}
For the spectral study, we consider data within the energy range 10 GeV--1 TeV. We calculate the SED of \hesssource\ using the best model obtained for the morphological study as discussed in Sec. \ref{sec:fermi_morphology}. The SED of \hesssource\ is shown in Fig. \ref{fig:sed_magic}, which is obtained by a fit to the data with the PL model.

 The best-fit PL model parameters are: prefactor, $N_0=(1.71 \pm 0.41) \times 10^{-14} ~\rm MeV^{-1} ~\rm cm^{-2}~\rm s^{-1}$, spectral index, $\alpha$ = $2.30  \pm 0.03$ and scale, $E_0$ = 1 GeV, where the uncertainties are statistical only. The total flux is found to be $F(>10 ~\rm GeV) = (1.2 \pm 0.1) \times 10^{-8} ~\rm photons ~\rm cm^{-2} ~\rm s^{-1}$.

%++++++++++++++++++++++++++++++++++++++++
\subsection{Joint fit to MAGIC and \textit{Fermi}-LAT data} \label{sec:joint-fit}
We perform a joint likelihood fit to the observed fluxes from MAGIC and \textit{Fermi}-LAT to find out the spectral behaviour of the source in the GeV--TeV energy range. We perform a $\chi^2$ fit on the \textit{Fermi}-LAT-MAGIC spectral points. We consider different spectral shapes as a PL, a PL with exponential cutoff (PLE) and a broken powerlaw (BPL) as spectral shapes for the fit. The PL has already been defined in subsection \ref{sec:magic_results_spectrum}. The PLE  spectral shape is defined as:
$$\rm PLE: {dN \over dE }= N_0 \left({E \over E_0}\right)^{-\alpha} \exp{(-{E \over E_0})},$$

The BPL model is defined as follows:
$$
\rm BPL:{\rm dN \over dE} = {
                     \begin{array}{ll}
                       A (E / E_0) ^ {-\alpha_1} & : E < E_{break} \\
                       A (E_{break}/E_0) ^ {\alpha_2-\alpha_1}
                           (E / E_0) ^ {-\alpha_2} & :  E > E_{break}, \\
                     \end{array}
                   }
$$

\noindent where $A$, $\alpha_1$, $\alpha_2$ and  $\rm E_{break}$ are parameters of the model. In the  case of the BPL model, the spectral break is at 37 GeV, while the cutoff energy in the PLE model is located at 1.8 TeV. Both the BPL and PLE models describe the SED better than a simple PL model, implying that a significant curvature is present in the SED. However, both BPL and PLE models show similar fit probability (p-value), making it difficult to favor one of them the most. {A combined fit to the SED with both BPL and PLE is shown in Fig. \ref{fig:sed_magic}.} The parameters of the different models, tested with $\chi^2$/d.o.f., are given in Table \ref{tab:joint-fit}.

\begin{table*}
    \centering
    \caption{Best-fit model parameters for the joint-fit to MAGIC and \textit{Fermi}-LAT spectral data points for three different models. The maximum-likelihood method is used to perform the joint-fit.} \label{tab:joint-fit}
    \begin{tabular}{c|c|c|c|c|c|c|c}
    \hline
    \hline
           model & amplitude & index1 & index2 & \rm e$_{cutoff}$ & \rm e$_{break}$ & $\chi^2$/d.o.f. & p-value\\
                 & $(\times 10^{-7})$ &  &       & (TeV)      & (TeV)       &           \\
           \hline
           PL    & 2.65 $\pm$ 0.21 & 2.23 $\pm$ 0.02 & -- & -- &  -- &196.4/10 & $9.1\times 10^{-37}$\\
           PLE   & 1.66 $\pm$ 0.21 & 1.92 $\pm$ 0.05 & -- & 1.8 $\pm$ 0.2 & --  & 27.6/9 & $1.1\times 10^{-3}$\\
           BPL   & 1.53 $\pm$ 0.22 & 1.91 $\pm$ 0.05 & 2.75 $\pm$ 0.10 & -- & 0.037 $\pm$ 0.005 & 20.8/8 & $7.8\times 10^{-3}$\\
    \hline 
    \hline
    \end{tabular}

    \label{tab:models}
\end{table*}

%%%%%%%%%%%%%%%%%%%%%%%%%%%%%%%%%%%%%%%%%
%%%%%%%%%%%%%%%%%%%%%%%%%%%%%%%%%%%%%%%%%
\section{Potential counterparts}\label{sec:conterparts}

Several point-like sources lie in the FoV of the extended gamma-ray source \hesssource\ and are likely to contribute to the VHE emission. In this section we discuss all these sources and their association with the observed emission. We consider some of the brightest emissions from these sources (see Fig. \ref{fig:gev-tev-associations}) discussed below to constrain the gamma-ray emission mechanisms in Section \ref{sec:Modelling}.

%+++++++++++++++++++++++++++++++
%+++++++++++++++++++++++++++++++
\subsection{G26.6-0.1}
\begin{figure}
\includegraphics[width=0.48\textwidth]{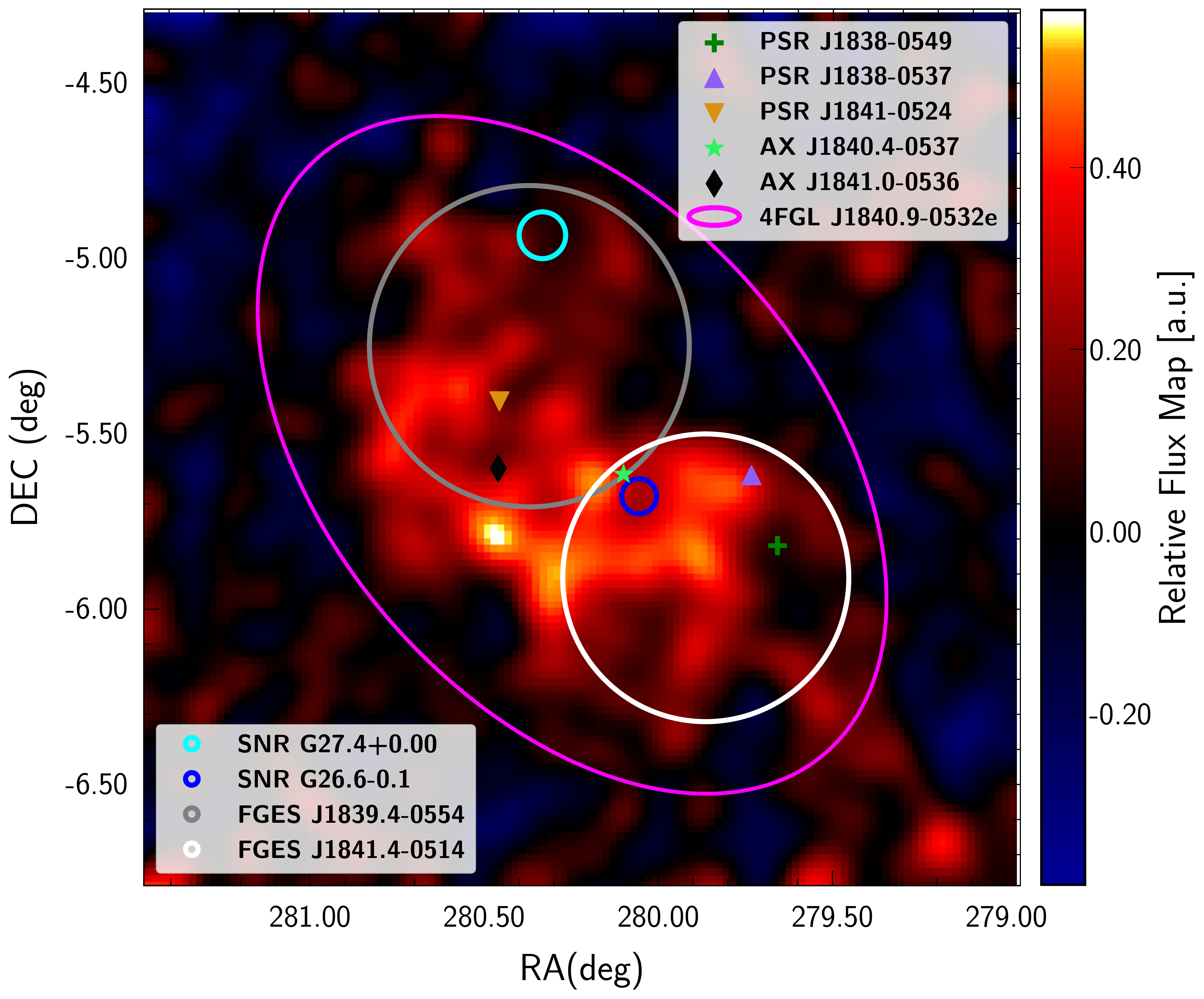} 
\caption{HE ($>$ 1 TeV) TS map as seen by MAGIC. Two extended Fermi-LAT sources FGES J1839.4--0554 and FGES J1841.4--0514 are overplotted (white and grey circles). The extension of the source reported in 4FGL catalog is shown as a magenta ellipse. Other point-like sources present in this region are also displayed (different markers).}\label{fig:gev-tev-associations}
\end{figure}

The diffuse hard X-ray source G26.6-0.1 was detected by ASCA in a Galactic plane survey \citep{Bamba_2003ApJ} which is located in this region as shown in Fig. \ref{fig:gev-tev-associations}. The observed X-ray spectrum was found to be featureless and can be fitted with a powerlaw function with photon index 1.3. The diffuse X-ray flux was estimated to be $\rm 3.5 \times 10^{-12} ~erg~ cm^{-2}~s^{-1}$ from a radius of 12$\arcmin$ region in the energy range of 0.7--7.0 keV. We consider this diffuse emission to be associated with \hesssource\ and assumed a corresponding scaled X-ray flux from a region with radius 0.4$^\circ$ similar to the extension of our source, for the multi-wavelength (MWL) modelling in Section \ref{sec:Modelling}. The distance to G26.6-0.1 is 1.3 kpc \citep{Bamba_2003ApJ}. Following this, the distance of \hesssource\ is assumed to be 2 kpc.

%+++++++++++++++++++++++++++++++
%+++++++++++++++++++++++++++++++
\subsection{PSR J1838--0537, PSR J1841-0524 and PSR J1838-0549 }
Several gamma-ray pulsars lie within the \hesssource\ region. \citet{Pletsch_2012ApJb} discovered the gamma-ray pulsar PSR J1838--0537 in a blind search of \textit{Fermi}-LAT data. It has been proposed as a potential candidate for the VHE source. It is a radio quiet pulsar. Also, no X-ray pulsation is observed from the location of the pulsar. If it is associated with a nebula, the subsequent observation from this region should have provided a detectable level of radio and X-ray fluxes from this region. The spin down power of the pulsar is estimated to be $\dot E= 5.9 \times 10^{36} \mathrm{erg~ s^{-1}}$. The integral energy flux of \hesssource\ estimated by MAGIC
over the range 0.1-10 TeV is $l{\gamma} \sim 9.13 \times 10^{-11} \mathrm{erg~ cm^{-2}~ s^{-1}}$. The luminosity for a distance of 2 kpc is, $L_{\gamma}=4\pi d^2 l{\gamma}$= $4.37 \times 10^{34} \mathrm{erg~s^{-1}}$ for isotropic emission. This implies
a conversion efficiency $\eta = L_{\gamma} /\dot E \sim 0.7$\% which is consistent with other suggested pulsar/pulsar wind nebulae (PWN) associations \citep{Hessels_2008ApJ}. Hence,
the pulsar's energetic is likely to power a PWN producing part of the TeV emission. The spectral index derived $\sim$2.4 is relatively soft in comparison to other PWNe detected at GeV energies by \textit{Fermi}-LAT.
Hence, part of the low-energy emission could have a different origin. 

There are two other known pulsars PSR J1841-0524 and PSR J1838-0549 \citep{Aharonian_2008_AA} which can contribute to the emission of \hesssource. The estimated  $\dot{E}/D^2$ values are given as $\rm 4.4 \times 10^{33}~ erg~ s^{-1} ~kpc^{-2}$ and $\rm 4.7 \times 10^{33}~ erg ~s^{-1}~ kpc^{-2}$ respectively and they can contribute to the observed emission when considered together. However, if taken separately, each would require approximately 200\% efficiency to explain the VHE emission \citep{Aharonian_2008_AA}. There is no significant radio emission observed from the location of these pulsars. The observed X-ray emission from these sources is lower than that considered for the multi-waveband modelling. Hence, assuming these constrains, we will not consider them separately for the MWL modelling. 

Hence, out of the three pulsars located in the region, it is most likely that only PSR J1838-0537 is contributing to the detected gamma-ray emission.

%+++++++++++++++++++++++++++++++
%+++++++++++++++++++++++++++++++
\subsection{FGES J1839.4--0554 \& FGES J1841.4--0514}
Recent \textit{Fermi}-LAT catalog for Galactic extended sources (FGESs) shows that there are two distinct extended sources in this region with energies above 10 GeV \citep{Ackermann_FGES_2017ApJ}. These sources, FGES J1839.4--0554 and FGES J1841.4--0514, are located at RA, {Dec} = 280$^\circ$.31 $\pm$ 0$^\circ$.04, -5$^\circ$.22 $\pm$ 0$^\circ$.03 and 279$^\circ$.90 $\pm$ 0$^\circ$.03, -5$^\circ$.90 $\pm$ 0$^\circ$.03 with an extension of 0$^\circ$.25 $\pm$ 0$^\circ$.02 and 0$^\circ$.31 $\pm$ 0$^\circ$.03, respectively. The extensions of these two sources are shown in Fig.  \ref{fig:gev-tev-associations}. It is evident from the figure that the observed GeV emission is overlapping with the extension found at TeV energies. Therefore, they can be considered as potential counterparts for the TeV emission. Although it appears that there are two different sources, the spectral characteristics at energies above 10 GeV are similar indicating that they may have common origin \citep{Ackermann_FGES_2017ApJ}. In our analysis of \textit{Fermi}-LAT data in this paper, we consider the entire region which includes both these sources to estimate the SED and we use it for the multi-wavelength modelling.

%+++++++++++++++++++++++++++++++
%+++++++++++++++++++++++++++++++
\subsection{G27.4+0.00 (Kes 73)}
  One of the sources studied in radio is a shell-type remnant Kes 73 (G027.4+00.0) which is present at the north-east of the extended emission region (see Fig. \ref{fig:gev-tev-associations}).  The small diameter $5\arcmin$ radio shell is characterized by a steep {spectral index ($\alpha \sim -0.68$, defined by $S \propto \nu^{\alpha}$)} between 0.5 to 5 GHz and flux density of 3.5 $\pm$ 0.5 Jy at 1.4 GHz \citep{Caswell1982MNRAS}. Radio studies of the remnant also show an incomplete shell structure with no {central engine} of Kes 73 {and} with a radio upper limits  of  0.45 mJy and  0.60 mJy at 6 cm and 20 cm radio wavelengths respectively \citep{Kriss1985ApJ}. This is considered to be unlikely counterparts due to the very small angular size of 5$'$ and its location on the edge of the extended emission. 

\subsection{AX J1840.4--0537 \& AX J1841.4--0536}
A weak point-like source, 1RXS J184049.1-054336, is located within G26.6-0.1 and it contributes to {less than} 10\% of the diffuse flux. Hence it is reasonable to exclude this weak X-ray flux from our analysis. The other X-ray point sources AX J1840.4-0537 and AX J1841.4-0536 are located outside the G26.6-0.1 region but well within the extended \hesssource. However the fluxes for these two sources were estimated to be $\rm 1.4 \times 10^{-13} ~erg~ cm^{-2}~s^{-1}$ and $\rm 2.1 \times 10^{-11} ~erg~ cm^{-2}~s^{-1}$, respectively, which are below the level of the scaled diffuse X-ray flux from this extended gamma-ray region. Moreover, due to the point-like morphology of these sources {with no associated nebula around them}, they can not be considered as potential counterparts of \hesssource. {However, a fraction of the total emission could be associated with these sources.}

Hence, although it is challenging to disentangle which sources are contributing to the observed GeV--TeV emission, we consider that the SNR G26.6-0.1, the pulsar PSR J1838-0537 and the extended FGES J1839.4-9554 and FGES J1841.4-0514 sources are the most promising counterparts, due to their energetics, extension and location within the region.

%%%%%%%%%%%%%%%%%%%%%%%%%%%%%%%%%%%%%%
%%%%%%%%%%%%%%%%%%%%%%%%%%%%%%%%%%%%%%
\section{Modelling of the spectral energy density}\label{sec:Modelling}
As already discussed above, there are several sources present in this extended region. Some of them are already argued to be potential counterparts at lower energies from the aspects of energetics of the system, while others are excluded due to their very point-like signatures along with the energetics {which can not contribute significantly to the overall emission}, considering the extent of the emission. In order to investigate if the multi-wavelength data can be explained self-consistently, we consider that the observed emission is associated with \hesssource. Since the radio and X-ray fluxes from the entire region of the extended emission can not be more than that estimated from different observations, we consider those results as upper-limits in these frequencies after multiplying with a scaling factor attributed to the extended region and the emission regions from where corresponding radio and X-ray measurements are performed. For the GeV-TeV modelling, we will consider the \textit{Fermi}-LAT and MAGIC data sets from this study and the H.E.S.S. data points from \cite{Aharonian_2008_AA}. We use the numerical code developed by \citet{Saha_2015} for the modelling.

%+++++++++++++++++++++++++++++++
%+++++++++++++++++++++++++++++++
\subsection{Leptonic model}\label{sec:leptonic}
  In Section \ref{sec:joint-fit}, we found that the SED has a spectral curvature and can be better explained with {either a BPL model or a PLE model}. Since the observed gamma-ray spectra carry imprints of the intrinsic particle distribution, a single population of electrons that follows a BPL type of distribution of electrons is assumed to calculate the Inverse Compton (IC) and bremsstrahlung emission processes.

In general, the electron spectrum might be more complicated than assuming a single population of electrons. For example, for the Crab Nebula, two different population of electrons are considered, namely, radio electrons and wind electrons. Radio electrons are less energetic electrons which reside in the nebular volume throughout its age, and they are mostly responsible for the observed radio fluxes. On the other hand, wind electrons are freshly accelerated electrons and they account for the observed fluxes at X-ray and GeV--TeV energies. {However, for simplicity, we consider a single population of electrons that is responsible for the observed emission at GeV--TeV energies.}

\begin{figure}
\includegraphics[width=0.5\textwidth]{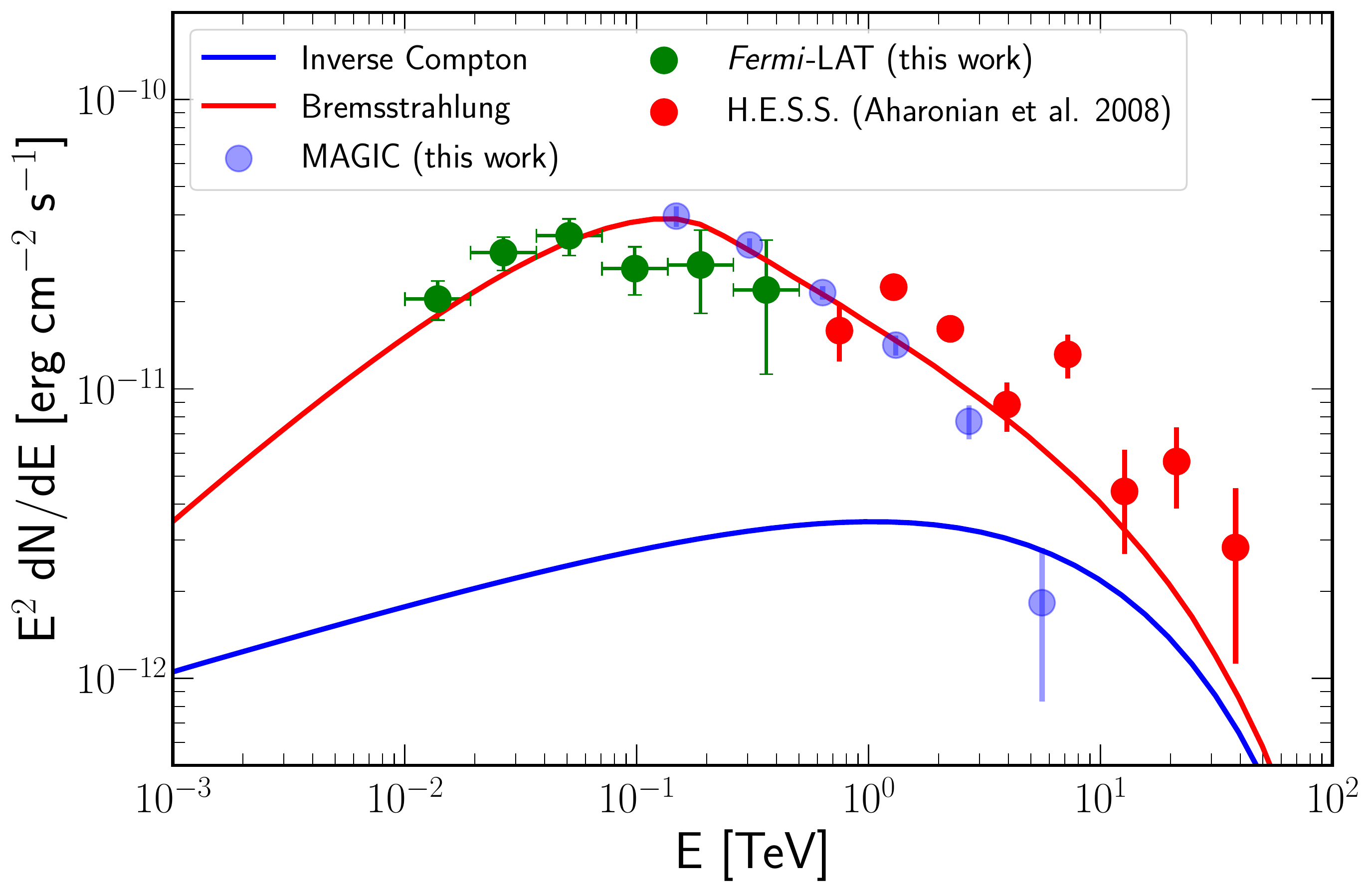}
\caption{The SED of \hesssource\ fitted with a leptonic model where IC (blue line) and bremsstrahlung (red line) emissions are considered to account for fluxes at GeV--TeV energies, \textit{Fermi}-LAT in green (this work), MAGIC in blue (this work) and H.E.S.S. in red \citep{Aharonian_2008_AA}. Bremsstrahlung emission spectrum, estimated for an ambient matter density of 100 cm$^{-3}$, is the dominant one. The parameters of the BPL electron distribution are shown in Table \ref{tab:fit_parameters}.} \label{fig:leptonic}
\end{figure}

We first consider that the observed gamma-ray radiation at GeV--TeV energies, is resulting from emission from relativistic electrons through IC and non-thermal bremsstrahlung processes. {For IC process, we consider that the high energy photons are produced by the up-scattering of photons from the Cosmic Microwave Background (CMB) and from {interstellar} dust contribution \citep{Mathis_1983}}. {For bremsstrahlung process}, we consider ambient matter density of 100 $\rm cm^{-3}$ following the estimation discussed in Appendix \ref{app:gas_density}. Higher or lower values of the ambient matter densities simply scale the contribution of bremsstrahlung spectrum. Fig. \ref{fig:leptonic} shows both IC and bremsstrahlung spectra for the BPL electron distribution for an ambient matter density of 100 $\rm cm^{-3}$.
The figure shows that the bremsstrahlung spectrum can explain the observed SED at GeV--TeV energies. On the other hand, the IC emission for the target photons of CMB and star lights can not explain the observed SED for the same population of electrons. The parameters of the BPL electron distribution are shown in Table \ref{tab:fit_parameters}. It is to be noted that the electron distribution can be adjusted to explain the observed emission by the IC spectrum. However, the bremsstrahlung spectrum will overestimate the observed flux for the same population of electron due to high ambient matter density. Hence bremsstrahlung becomes dominant emission process within leptonic scenario.

\begin{figure*}
\centering
\includegraphics[width=\textwidth]{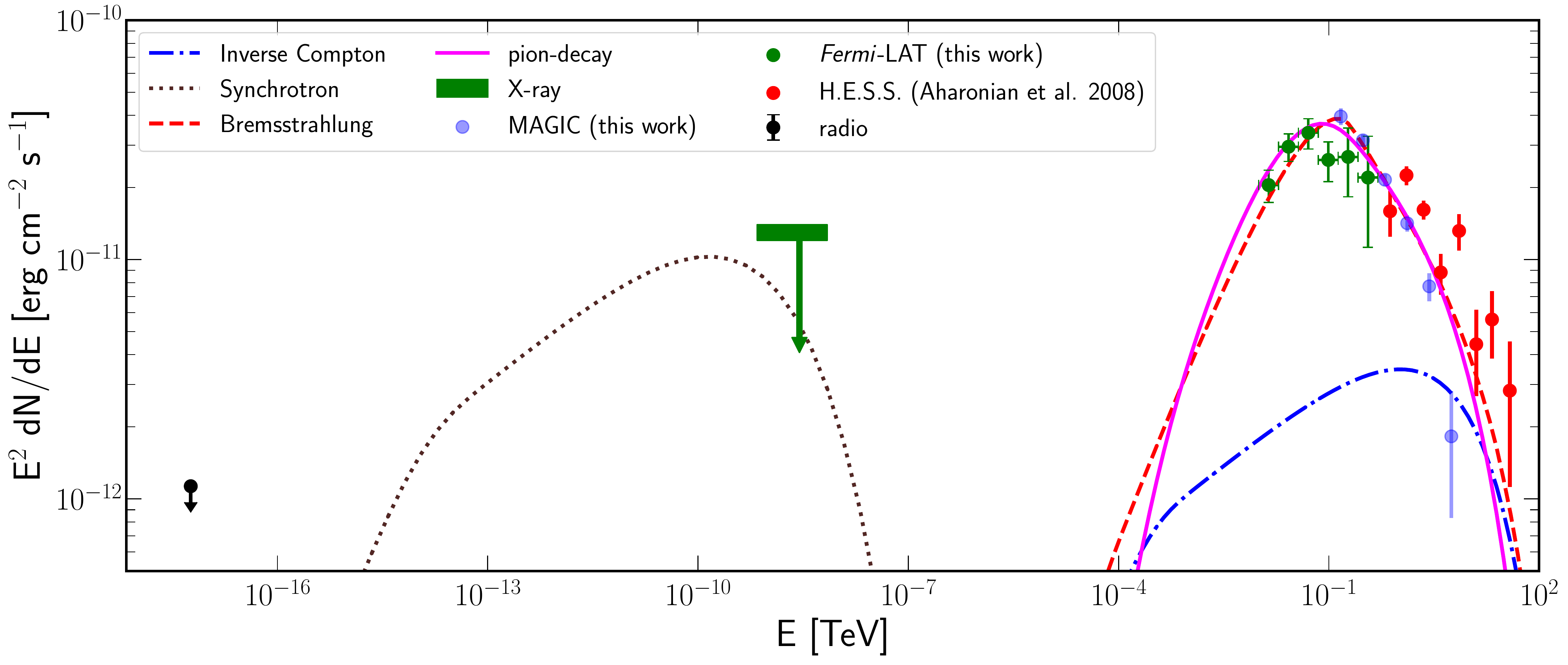}
\caption{SED of \hesssource\ from radio to TeV energies. HE \textit{Fermi}-LAT data (green points) and TeV MAGIC (blue points) and H.E.S.S. (red points) data are fitted both with hadronic (magenta solid line) and bremsstrahlung (red dashed line) models. Inverse Compton (blue dot-dashed line) can not account for the measured gamma-ray emission. The synchrotron spectrum (brown dotted line) is fitted according to the radio and X-ray ULs.}
\label{fig:sed_multi}
\end{figure*}

In order to check the contribution of the synchrotron spectrum for the electron distribution, we calculate the synchrotron spectrum leaving magnetic field as a  free parameter. We find that the synchrotron spectrum for a magnetic field of approximately  5 $\mu$G does not overestimate the radio and X-ray limits estimated for this study. The synchrotron component {only} contributes to {radio and X-ray energies}. The synchrotron spectrum together with IC and bremsstrahlung spectra is shown in Fig. \ref{fig:sed_multi}.

%+++++++++++++++++++++++++++++++
%+++++++++++++++++++++++++++++++
\subsection{Hadronic model}\label{sec:hadronic}

\begin{figure}
\includegraphics[width=0.5\textwidth]{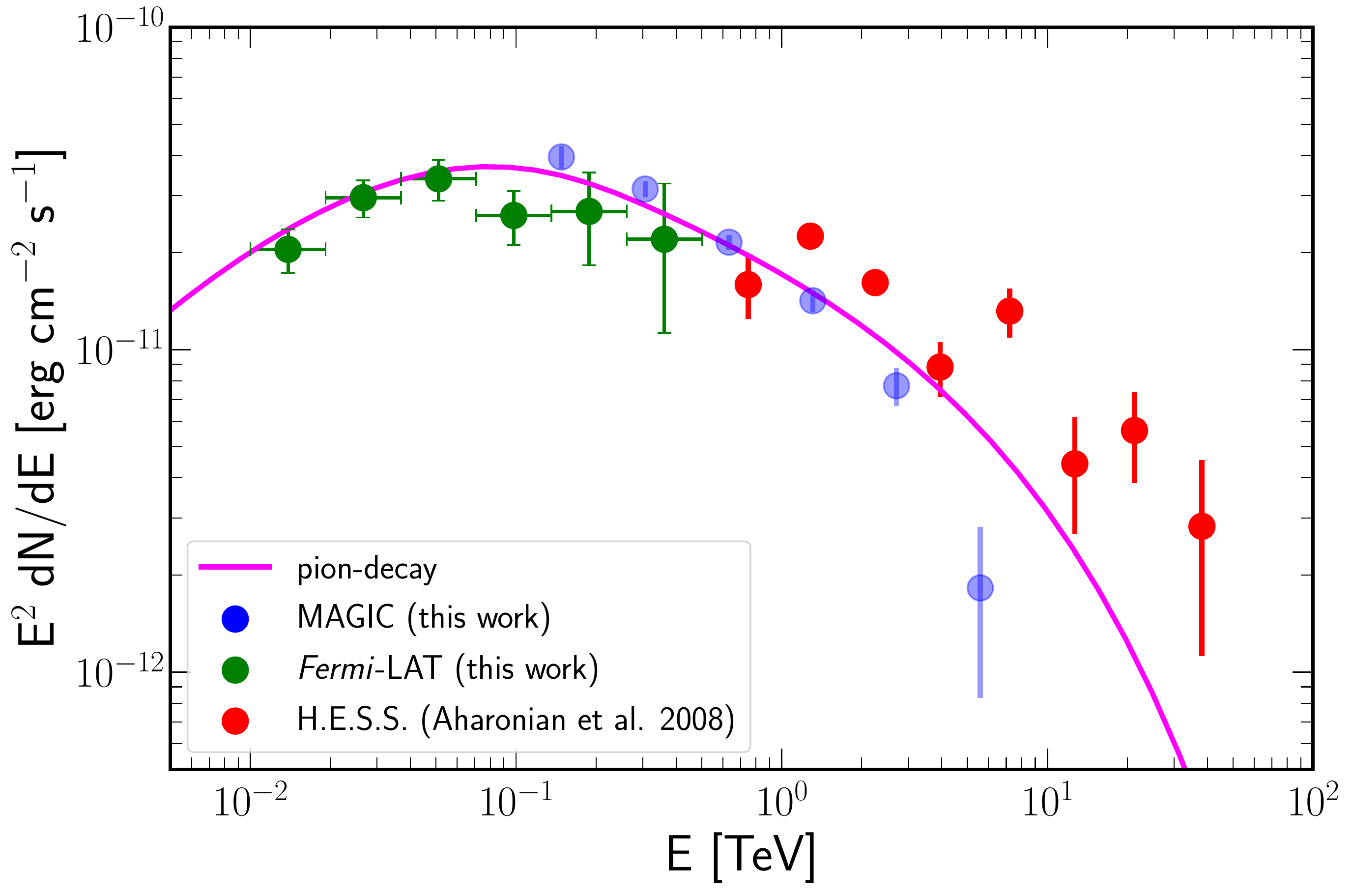}
\caption{SED of \hesssource, fitted with a $\pi^0$-decay emission spectrum, assuming the parameters shown in Table \ref{tab:fit_parameters}.} \label{fig:hadronic}
\end{figure}

We also introduce a hadronic scenario as an additional component which contributes significantly at {gamma rays} (GeV--TeV). We calculate the gamma-ray spectrum resulting from the decay of neutral pions  following \citealt{Kelner_2006}. The gamma-ray spectrum for the relativistic protons for the BPL model as considered for the leptonic model and for  an ambient gas density of $\rm n_0$ $\simeq 100$ cm$^{-3}$ is shown in Fig. \ref{fig:hadronic}. The total energy  can be calculated  as $\rm W_p =  5.52 \times 10^{48} \times (100.0/\rm n_0)$  erg. The figure displays that the gamma-ray spectrum resulting from the decay of neutral pions can explain the observed GeV-TeV data very well. The parameters of the model are presented in Table \ref{tab:fit_parameters}.

\begin{table}
\caption{Parameters for physical models for a single zone particle distribution of a BPL model. The parameters are obtained considering two different models: leptonic and hadronic. Parameters with errors are used as free parameters for the fit.}
\label{tab:fit_parameters}
\centering
\begin{tabular}{c|c|c}
\hline
\hline
Parameters &     Leptonic & Hadronic  \\
           &              &           \\
\hline 
& &\\
Spectral index  ($\alpha_1$)                 &  1.06$_{-0.05}^{+0.10} $              & 1.18$_{-0.11}^{+0.13}$ \\
 & &\\
Spectral index  ($\alpha_2$)                 &  2.52$_{-0.05}^{+0.06}$             & 2.02$_{-0.05}^{+0.05}$ \\
& &\\
Energy at spectral break, \rm E$_b$ \rm (TeV)        &  0.18$_{-0.02}^{+0.03}$             & 0.22$_{-0.04}^{+0.07}$ \\
& &\\
Ambient matter density, $n_0$ ($\rm cm^{-3}$)  &   100                 & 100\\
& &\\
Total energy      ($10^{48}$ erg)         &  5.82$_{-0.19}^{+0.11}$                & 5.52$_{-0.13}^{+0.12}$  \\
& &\\
\hline
\hline
\end{tabular}
\end{table}

%%%%%%%%%%%%%%%%%%%%%%%%%%%%%%%%%%%%%%%%%%
%%%%%%%%%%%%%%%%%%%%%%%%%%%%%%%%%%%%%%%%%%
\section{Discussion}\label{sec:Discussion}
%+++++++++++++++++++++++++++++++
%+++++++++++++++++++++++++++++++
\subsection{Gamma rays from extended unidentified sources}
The analysis of about 34 hrs of good quality MAGIC data confirms that the gamma-ray emission is as extended as claimed by the H.E.S.S. Collaboration and the source is detected with high significance for energy above 50 GeV. In addition to that, we investigate the source morphology as a function of energy. The observed results suggest that at low energies the overall region is detected like a diffuse source, with some few regions around the center of the source where the significance is higher than 5$\sigma$. This indicates the possibility that several point-like sources are contributing to the extended emission. At medium energies, between 500 GeV -- 1 TeV, the emission is concentrated along the center, in a North-South line. In addition to that, the skymap above 1 TeV shows that the extension of the emission is reduced compared to that of low energies and the most significant flux is located at the southern part of the extended region, with only few hot spots over 5$\sigma$. {The morphological analysis of MAGIC data also shows that the multiple source model is better over a single-source model. This establishes the fact that several sources are contributing to the extended emission.}

The morphological study of the source using about 10 years of \textit{Fermi}-LAT data above 10 GeV also shows that the source is extended.  However, the spectral shape is different from that of MAGIC. When comparing with the emission at higher (TeV) energies, \hesssource\ displays an extension compatible to that measured by MAGIC.
%is as extended \textbf{as} measured by MAGIC.

In the case of the \textit{Fermi}-LAT detection, the spectrum of the source is best described by a powerlaw. To study the spectral behavior within the entire energy range, from {GeV} to TeV energies, we performed a joint fit on the spectral data points from MAGIC and \textit{Fermi-LAT}. We found that the combined SED is best described either by a broken powerlaw model with a spectral break at $\sim$37 GeV or with a powerlaw with exponential cut-off {at 1.8 TeV}.

%+++++++++++++++++++++++++++++++
%+++++++++++++++++++++++++++++++
\subsection{Emission mechanisms}
Multi-wavelength modelling of the data indicates that the leptonic model can explain the data well. Due to the higher ambient matter density, the bremsstrahlung spectrum dominates over IC spectrum. The radio and X-ray fluxes put a constraint on the magnetic field in the emission volume when they are accounted with a synchrotron emission process. The magnetic field  of 5 $\mu$G as mentioned in section \ref{sec:leptonic} is very close to that of some other known old PWNe \citep{Reynolds_2012,Kargaltsev_2013}. Given the high ambient matter density and presence of molecular clouds, a  hadronic emission model is also suitable to explain the observed data at GeV--TeV energies. We find that the hadronic model can explain the data very well for a BPL proton distribution and an ambient matter density of $100/\rm cm^3$. Therefore, both leptonic and hadronic model can explain the data well with the parameters shown in Table \ref{tab:fit_parameters}.

In the whole discussion on multi-wavelength modelling of the data, our assumption was that the observed emission is entirely due to a single source. However, we have already seen that the region is populated by different sources which were established through observations at lower energy bands. Some of the sources are already excluded to be considered as potential gamma-ray emitters while energetic is considered. However, some of them could be potentially associated with the observed emission at GeV--TeV energies. Given the angular resolution of the gamma-ray telescopes at present generation, it is not possible to have an unambiguous association with the sources at other wavebands. One possible scenario for the extension of the emission is the interaction of run-away cosmic particles from the source and the gamma-ray visibility is enhanced due to interaction with molecular clouds which are covering the extended source very well as can be seen from Fig. \ref{fig:skymaps_with_MC} and discussed in appendix A. The presence of molecular clouds along the extension of the source also supports the relatively high ambient matter density required for both leptonic and hadronic model.

\begin{figure}
\includegraphics[scale=0.27]{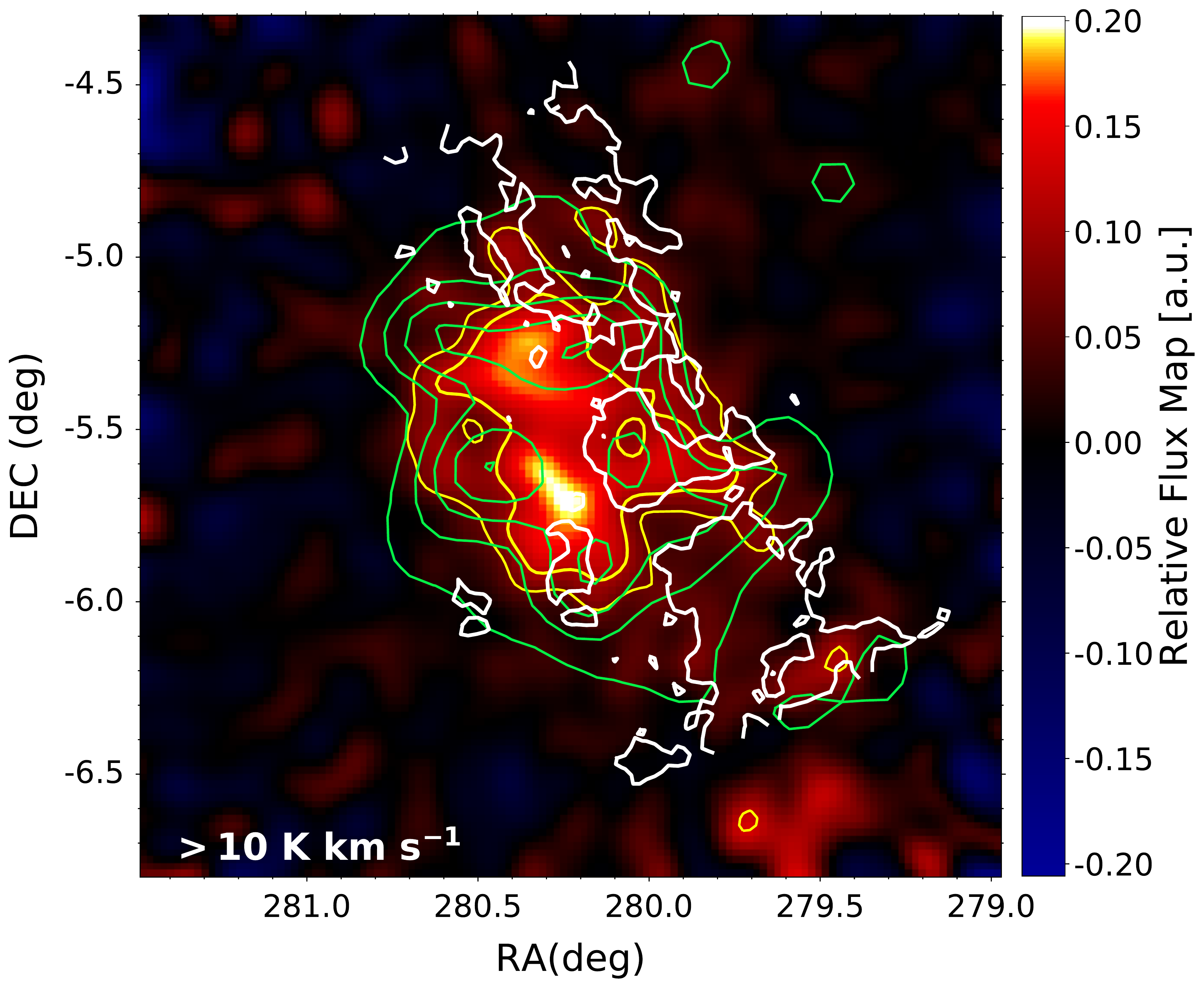}
\caption{LE skymap, similar to Fig. \ref{fig:skymaps}. The source as seen by MAGIC is shown as yellow contours, while the \textit{Fermi}-LAT source is plotted as green contours. The CO contours are shown with white solid lines when CO map is integrated over the range of -5 km/s to 135 km/s. The CO data is obtained from archival Galactic Ring Survey \citep[GRS;][]{GRS_data_2006ApJS}. }\label{fig:skymaps_with_MC}
\end{figure}

%++++++++++++++++++++++++++++++++++++++
%++++++++++++++++++++++++++++++++++++++
\subsection{The nature of \hesssource}\label{sec:associations}
The observations at X-ray energies did not show any bright synchrotron nebula around the pulsars present in this region.  However, in this scenario the absence of bright synchrotron nebula can be  easily explained. If the TeV source is powered by one or several pulsars present in this region, then pulsars are expected to be relic ones. For such PWNe, IC emission efficiency is more pronounced due to lower magnetic fields. In Section \ref{sec:Modelling}, we find that the IC contribution for this source is 10\% less compared to the bremsstrahlung spectrum. Therefore, it is reasonable to consider that the synchrotron emission could be even more inefficient, which supports the absence of the bright synchrotron nebula around the pulsars. Therefore, if the bright TeV emission is assumed to be associated with a PWN, then the PWN requires to be a relic one where the remnant of the supernova explosion has already disappeared. 

It is also discussed in Section \ref{sec:conterparts} that energetically the pulsar PSR J1838-0537 is able to account for the observed GeV--TeV energies. The gamma-ray flux at TeV energies, a factor two lower than the Crab nebula flux, is required to have, $S_0$ = ($\rm L_0/10^{37} \rm erg~ s^{-1})(d/1 ~kpc)^{-2} \geq 10^{-3}$, where $\rm L_0$ and $d$ are the luminosity and distance of the source respectively \citep{Aharonian_book_2004}.  The two known pulsars PSR J1841-0524 and PSR J1838-0549 cannot contribute to the observed GeV--TeV energies since $S_0$ is less than $10^{-3}$. However, $S_{0}$ is greater than ~$10^{-3}$ for PSR J1838-0537 making it a potential counterpart of \hesssource. Since PSR J1838-0537 is not a part of any radio or X-ray nebula, it is also possible to consider that it is an isolated pulsar which has already left the remnant. TeV emission is an effective product of the IC mechanism for such isolated pulsars, with the injection of relativistic electrons in the interstellar magnetic field which is about 3$\mu$G. In such a scenario, the bright X-ray and radio synchrotron nebula could be absent.  However, the extension of such a source is not readily accepted. Nevertheless, the presence of the molecular clouds along the observed GeV--TeV emission can support its extension within this scenario {but through bremsstrahlung processes}.

The extension of the source is estimated to be 0.4$^\circ$ when the source is fitted with a 2D Gaussian with a equal spatial width for both the directions. This extension translates into a radius of approximately 14 pc at a source distance of 2 kpc. The effective diffusion radius can be calculate as , $R_{diff} \simeq 2 \sqrt{D(E)~t}$, where $D(E)$ is the diffusion coefficient and can be represented as $D(E) = D_0 (E/10 GeV)^\delta$ \citep{AtoyanPhysRevD.52.3265}. {The commonly used diffusion coefficient at 10 GeV is of about $D_0 \sim 10^{28} \rm cm^{2} \rm s^{-1}$} \citep{Ginzburg:1990sk} and assuming that $\delta=0.5$ ($\delta$ is one of the parameters of the diffusion coefficients; for energy independent diffusion coefficient $\delta$=0), we calculate the diffusion time scale of {$t_{diff}=17$ kyr}. On the other hand, the lifetime of the bremsstrahlung loss, which is independent of energy, is estimated as $t_{brems} \simeq 4 \times 10^{4}~(n/1~\rm cm^{-3})^{-1}$ kyr $=4 \times 10^{2}$ kyr for ambient matter density of 100 $\rm cm^{-3}$. Therefore, the dominant emission through bremsstrahlung process for the estimated ambient matter density is a viable solution for the observed extension of the source. We then conclude that the observed emission can be potentially associated with a PWN.

The observed emission can also be considered to be associated with SNRs, since it is seen that there are two SNRs present in and around the source. The first one, G27.4+0.00, is located at the edge of the TeV emission and has a relatively small angular size, hence is unlikely that it can account for the observed gamma rays. G26.6-0.1 is considered to be a potential counterpart for the extended emission. However, there are no strong radio and X-ray nebulae associated with the extent of the emission which is the case for a typical SNR scenario. Hence, a strong association can be made provided that the observed emission is considered due to the particles that escaped the SNR shocks and are interacting with the molecular clouds. Following the diffusion timescale as discussed in the preceding paragraph and the age of the SNR G26.6-0.1 as a middle-age SNR \citep[$\sim$10$^3$ years;][]{Bamba_2003ApJ}, it can be considered a possible candidate for at least part of the detected GeV-TeV emission.

%%%%%%%%%%%%%%%%%%%%%%%%%%%%%%%%%%%%%%%%
%%%%%%%%%%%%%%%%%%%%%%%%%%%%%%%%%%%%%%%%
\section{Summary and Conclusions} \label{sec:conclusion}
We report a deep study of the unidentified gamma-ray source \hesssource\ at GeV--TeV energies using about 34 hours of MAGIC and 10 years of \textit{Fermi}-LAT data. We summarize the results below.
\begin{itemize}
    \item The results of the detailed analysis show that the observed gamma-ray emission from \hesssource\ is significantly extended. The estimated extension of the source using MAGIC data is similar to that reported by the H.E.S.S. Collaboration, found to be $\sim$ 0.4$^\circ$ assuming a Gaussian distribution.
    \item There are several bright hot-spots in the extension of the source which appears to be multiple sources which contribute  to the observed emission at GeV-TeV energies. The emission at TeV energies moves towards the south with increasing energy, revealing this region as one of the potential main contributors of the TeV extended emission. 
    \item {The extended emission is modelled better with a multi-source model compared to a single-source model.}
    \item The spectral curvature of the SED in the energy range from GeV--TeV is significant and {it can either be described by a broken powerlaw model with break at 37 GeV or a powerlaw with exponential cutoff at 1.8 TeV}.
    
    \item The observed SED can be explained well with both a leptonic (bremsstrahlung) and a hadronic model for the density of ambient matter of $100~ {\rm cm^{-3}}$ {assuming} a BPL distribution of electrons and protons, respectively.
\end{itemize}

Within the present morphological and spectral studies of this extended source using GeV--TeV data and available MWL information on sources present within the region, we conclude that the extended gamma-ray emission seems to be associated with multiple sources in this region. The GeV--TeV emission is compatible with a PWN scenario, although a fraction of the gamma-ray emission can also be explained within a SNR scenario. However, disentangling these sources {at TeV energies (either point sources or extended sources)} from {one another} and quantifying their contribution to the observed morphology of the source demands much better angular resolution compared to the present generation of gamma-ray telescopes. Hence, it becomes naturally an interesting source of study for the next generation of IACT telescopes.

%%%%%%%%%%%%%%%%%%%%%%%%%%%%%%%%%%%%%%%%
%%%%%%%%%%%%%%%%%%%%%%%%%%%%%%%%%%%%%%%%
\begin{appendix}
\section{Target gas density}\label{app:gas_density}
Here we evaluate the target gas density required both for leptonic and hadronic models as discussed in Sec. \ref{sec:leptonic} and Sec. \ref{sec:hadronic}. To evaluate the target gas density, we estimate the densities of each gas phase (neutral hydrogen HI and molecular hydrogen H$_{2}$) and then sum the estimated values to get the total contribution to the gas density.
Under the assumption of the optically thin limit, the HI column density is given by \citep{Dickey1990}, $\rm N(HI) \simeq 1.823 \times 10^{18}\int T_b(HI; v_r)dv_r ~cm^{-2}$, where
$\rm T_b(HI;v_r)$ is the brightness temperature of the observed
21 cm line at $v_{r}$. In order not to overestimate the gas density within the source, we need to integrate over some range of $\rm v_r$. We consider
$\rm v_r$ in the range of 110 - 135 $\rm km ~ s^{-1}$ corresponding to the distance of about 2 kpc.
The average HI density is estimated to be $\rm N(HI) = 8.65 \times 10^{20} \rm cm^{-2}$ for a radius of 0.4$^\circ$ centered on the \hesssource\ using the database of the HI4$\pi$ survey \citep{HI4PI_2016}. We assume that the HI gas  is uniformly distributed within the source. The length of the line of sight along the direction of the \hesssource\ is $\ell= 2r_{0}$, where $\rm r_0$ is the radius of the extended emission. The radius of the source is considered to be $0^\circ.4$ which translates to approximately 14 pc for a distance of 2 kpc to the source. The density of the neutral hydrogen gas is $\rm n(HI) = \rm N(HI)/\ell \simeq~ 10~ \rm cm^{-3}$. 
We use observations (see Fig. \ref{fig:skymaps_with_MC}) of the $\rm ^{13}CO (J=1-0)$ line, which traces molecular clouds, from the Galactic Ring Survey \citep[GRS;][]{GRS_data_2006ApJS}.The CO spectrum over the range of velocities from +110 to +135 km $\rm s^{-1}$ are integrated to obtain the velocity integrated CO intensity ($\rm W_{CO}$). The $\rm W_{CO}$ averaged
over the region with a radius of 0$^\circ$.4 covering the extended emission is found to be approximately 63 K km $\rm s^{-1}$. To estimate the mass of the molecular cloud, we use the standard CO-to-H2 conversion factor of $\rm X_{CO} = N(H2)/WCO = 1.8 \times 10^{20} cm^{-2} K^{-1} km^{-1}~ s$ \citep{Dame2001ApJ}. 
We find N(H2) = $4.8 \times 10^{22} \rm cm^{-2}$. Therefore, the density of the molecular hydrogen gas, $\rm n(H_2) = \rm N(H_2)/\ell \simeq ~ 130~ \rm cm^{-3}$. The total gas density, hence, is $\rm n(HI) + \rm n(H_2) \simeq 140~ \rm cm^{-3}$. However, for simplicity we consider the gas density of 100 $\rm cm^{-3}$ for the physical modelling of the source.
\end{appendix}

%%%%%%%%%%%%%%%%%%%%%%%%%%%%%%%%%%%
%%%%%%%%%%%%%%%%%%%%%%%%%%%%%%%%%%%
\section*{Acknowledgments}
We would like to thank the Instituto de Astrofisica de Canarias for the excellent working conditions at the Observatorio del Roque de los Muchachos in La Palma. The financial support of the German BMBF and MPG, 
the Italian INFN and INAF, the Swiss National Fund SNF, the ERDF under the Spanish MINECO (FPA2015-69818-P, FPA2012-36668, FPA2015-68378-P, FPA2015-69210-C6-2-R, FPA2015-69210-C6-4-R, FPA2015-69210-C6-6-R, AYA2015-71042-P, AYA2016-76012-C3-1-P, ESP2015-71662-C2-2-P, CSD2009-00064), 
and the Japanese JSPS and MEXT is gratefully acknowledged. 
This work was also supported by the Spanish Centro de Excelencia "Severo Ochoa" SEV-2012-0234 and SEV-2015-0548, and Unidad de Excelencia "Maria de Maeztu" MDM-2014-0369, by the Croatian Science Foundation (HrZZ) Project 09/176 and the University of Rijeka Project 13.12.1.3.02, by the DFG Collaborative Research Centers SFB823/C4 and SFB876/C3, and by the Polish MNiSzW grant 2016/22/M/ST9/00382. 

This publication makes use of molecular line data from the Boston University-FCRAO Galactic Ring Survey (GRS). The GRS is a joint project of Boston University and Five College Radio Astronomy Observatory, funded by the National Science Foundation under grants AST-9800334, AST-0098562, AST-0100793, AST-0228993, \& AST-0507657.

\section*{Data availability}
Raw data were generated at the MAGIC telescopes large-scale facility. Derived data supporting the findings of this study are available from the corresponding authors upon request. Proprietary data reconstruction codes were also generated at the MAGIC telescope large-scale facility.  The \textit{Fermi-}LAT data are available in the public domain: \url{https://fermi.gsfc.nasa.gov/ssc/data/access/}. Information supporting the findings of this study is available from the corresponding authors upon request.

\bibliographystyle{mnras}
\bibliography{masterbibtex}

% Don't change these lines
\bsp	% typesetting comment
\label{lastpage}
\vspace{0.5cm}

$^{1}$ { Inst. de Astrof\'isica de Canarias, E-38200 La Laguna, and Universidad de La Laguna, Dpto. Astrof\'isica, E-38206 La Laguna, Tenerife, Spain
$^{2}$ Universit\`a di Udine, and INFN Trieste, I-33100 Udine, Italy
$^{3}$ National Institute for Astrophysics (INAF), I-00136 Rome, Italy
$^{4}$ ETH Zurich, CH-8093 Zurich, Switzerland
$^{5}$ Japanese MAGIC Consortium: ICRR, The University of Tokyo, 277-8582 Chiba, Japan; Department of Physics, Kyoto University, 606-8502 Kyoto, Japan; Tokai University, 259-1292 Kanagawa, Japan; Physics Program, Graduate School of Advanced Science and Engineering, Hiroshima University, 739-8526 Hiroshima, Japan; Konan University, 658-8501 Hyogo, Japan; RIKEN, 351-0198 Saitama, Japan
$^{6}$ Technische Universit\"at Dortmund, D-44221 Dortmund, Germany
$^{7}$ Croatian Consortium: University of Rijeka, Department of Physics, 51000 Rijeka; University of Split - FESB, 21000 Split; University of Zagreb - FER, 10000 Zagreb; University of Osijek, 31000 Osijek; Rudjer Boskovic Institute, 10000 Zagreb, Croatia
$^{8}$ IPARCOS Institute and EMFTEL Department, Universidad Complutense de Madrid, E-28040 Madrid, Spain
$^{9}$ Centro Brasileiro de Pesquisas F\'isicas (CBPF), 22290-180 URCA, Rio de Janeiro (RJ), Brasil
$^{10}$ University of Lodz, Faculty of Physics and Applied Informatics, Department of Astrophysics, 90-236 Lodz, Poland
$^{11}$ Universit\`a  di Siena and INFN Pisa, I-53100 Siena, Italy
$^{12}$ Deutsches Elektronen-Synchrotron (DESY), D-15738 Zeuthen, Germany
$^{13}$ Centro de Investigaciones Energ\'eticas, Medioambientales y Tecnol\'ogicas, E-28040 Madrid, Spain
$^{14}$ Istituto Nazionale Fisica Nucleare (INFN), 00044 Frascati (Roma) Italy
$^{15}$ Max-Planck-Institut f\"ur Physik, D-80805 M\"unchen, Germany
$^{16}$ Institut de F\'isica d'Altes Energies (IFAE), The Barcelona Institute of Science and Technology (BIST), E-08193 Bellaterra (Barcelona), Spain
$^{17}$ Universit\`a di Padova and INFN, I-35131 Padova, Italy
$^{18}$ Universit\`a di Pisa, and INFN Pisa, I-56126 Pisa, Italy
$^{19}$ Universitat de Barcelona, ICCUB, IEEC-UB, E-08028 Barcelona, Spain
$^{20}$ The Armenian Consortium: ICRANet-Armenia at NAS RA, A. Alikhanyan National Laboratory
$^{21}$ Universit\"at W\"urzburg, D-97074 W\"urzburg, Germany
$^{22}$ Finnish MAGIC Consortium: Finnish Centre of Astronomy with ESO (FINCA), University of Turku, FI-20014 Turku, Finland; Astronomy Research Unit, University of Oulu, FI-90014 Oulu, Finland
$^{23}$ Departament de F\'isica, and CERES-IEEC, Universitat Aut\`onoma de Barcelona, E-08193 Bellaterra, Spain
$^{24}$ Saha Institute of Nuclear Physics, HBNI, 1/AF Bidhannagar, Salt Lake, Sector-1, Kolkata 700064, India
$^{25}$ Inst. for Nucl. Research and Nucl. Energy, Bulgarian Academy of Sciences, BG-1784 Sofia, Bulgaria
$^{26}$ now at University of Innsbruck
$^{27}$ also at Port d'Informaci\'o Cient\'ifica (PIC) E-08193 Bellaterra (Barcelona) Spain
$^{28}$ also at Dipartimento di Fisica, Universit\`a di Trieste, I-34127 Trieste, Italy
$^{29}$ also at INAF-Trieste and Dept. of Physics \& Astronomy, University of Bologna
}      
\end{document}